\documentclass{myjfm}
\usepackage{natbib}

\usepackage{psfrag}
\usepackage{color}
\usepackage{graphicx,float,amssymb}
\usepackage[bf,loose,normalsize,FIGTOPCAP]{subfigure}
\newcommand{\p}{{\partial}}

\newcommand{\rd}{{\rm d}}
\newcommand{\rD}{{\rm D}}
\newcommand{\beq}{\begin{equation}}
\newcommand{\eeq}{\end{equation}}
\newcommand{\bqa}{\begin{eqnarray}}
\newcommand{\eqa}{\end{eqnarray}}

\newcommand{\lm}{{\lambda}}
\newcommand{\lmn}{{\lambda_{mn}}}
\newcommand{\kmn}{{k_{mn}}}
\newcommand{\rmn}{{\rho_{mn}}}

\renewcommand{\th}{{\theta}}

\newcommand{\re}{\mathrm{Re}}
\newcommand{\ra}{\mathrm{Ra}}

\newcommand{\Pra}{\mathrm{Pr}}
\newcommand{\Nu}{\mathrm{Nu}}

\newcommand{\wrms}{{w_{\rm{rms}}}}
\newcommand{\uhrms}{{u_{h,\rm{rms}}}}
\newcommand{\ti}{\times}
\title{Axially-homogeneous Rayleigh-B\'{e}nard convection in a cylindrical cell}

\author[L. E. Schmidt et al.]{LAURA E. SCHMIDT$^1$, ENRICO CALZAVARINI$^2$, DETLEF LOHSE$^1$, FEDERICO TOSCHI$^{3,4}$ and ROBERTO VERZICCO$^{1,5}$}

\affiliation{$^1$ Physics of Fluids, Department of Science and Technology, Impact and Mesa+ Institutes and 
J. M. Burgers Center for Fluid Dynamics, University of Twente, PO Box 217, 7500 AE Enschede, The Netherlands

$^2$ Laboratoire de M\'ecanique de Lille CNRS/UMR 8107, Universit\'e Lille 1  and  Polytech'Lille, Cit\'e Scientifique Av.~P.~Langevin, 59650 Villeneuve d'Ascq, France

$^3$ Department of Physics, and Department of Mathematics and Computer Science, and J.M. Burgers Center for Fluid Dynamics, Eindhoven University of Technology, 5600 MB  Eindhoven, The Netherlands 

$^4$ CNR-IAC, Via dei Taurini 19, 00185 Rome, Italy

$^5$Department of Mechanical Engineering, University of Rome ``Tor Vergata", via del Politecnico 1, 00133 Rome, Italy
}  \date{\today} 

\begin{document}
\maketitle

\begin{abstract}
Previous numerical studies have shown that the ``ultimate regime of thermal convection" can be attained in a Rayleigh-B\'{e}nard cell when the kinetic and thermal boundary layers are eliminated by replacing the walls with periodic boundary conditions (homogeneous Rayleigh-B\'{e}nard convection).  Then, the heat transfer scales like $\Nu\sim\ra^{1/2}$ and turbulence intensity as $\re\sim\ra^{1/2}$, where the Rayleigh number $\ra$ indicates the strength of the driving force. However, experiments never operate in unbounded domains and it is important to understand how {\it confinement} might alter the approach to this ultimate regime.  Here we consider 
homogeneous Rayleigh-B\'{e}nard convection in a laterally confined geometry -- a small aspect-ratio vertical cylindrical cell -- and show evidence of the ultimate regime as $\ra$ is increased:
In spite of the confinement and the resulting kinetic boundary layers, we still find 
$\Nu \sim \re \sim \ra^{1/2}$. 
The system supports exact solutions composed of modes of exponentially growing vertical velocity and temperature fields, 
with $\ra$ as the critical parameter determining the properties of these modes.  Counterintuitively, in the low $\ra$ regime, or for very narrow cylinders,  the numerical simulations are susceptible to these solutions which can dominate the dynamics and lead to very high and unsteady heat transfer.   As $\ra$ is increased,  interaction between modes stabilizes the system, evidenced by the increasing homogeneity and reduced fluctuations in the r.m.s. velocity and temperature fields.  We also test that physical results become independent of the periodicity length of the cylinder, a purely numerical parameter, as the aspect ratio is increased. 
\end{abstract}

\section{Introduction}
There has been longstanding scientific interest in the heat transfer in Rayleigh-B\'{e}nard convection (RBC) -- when a fluid confined between two plates undergoes thermal convection due to a temperature difference between the cold top and hotter bottom plate ~(\cite{kad01,sig94,ahl09,loh10}).  The dynamics of the system depend on the strength of the driving temperature difference (given by the Rayleigh number, $\ra$) and the ratio of the kinematic viscosity to the thermal diffusivity (given by the Prandtl number, $\Pra$). Recently, studies of unconfined thermal convection -- starting with the numerical work by \cite{loh03} --
 were prompted by the relevance to natural convection phenomena, as occurs in the Earth's atmosphere (\cite{cel07b}) or in the core of stars (\cite{gar10}), and as a way to test theoretical predictions for the scaling of the heat transfer (given by the Nusselt number, $\Nu$) and turbulence intensity (given by the Reynolds number, $\re$). 

In the limit of high $\ra$ it was proposed that the dynamics would reach an ultimate regime of thermal convection, which is dominated by the bulk flow rather than the viscous and thermal boundary layers~(\cite{kra62}). Within the Grossmann-Lohse theory of thermal convection, this regime where the dynamics does not depend on thermal plumes is also predicted~(\cite{gro00,gro01,gro02}).  Under these conditions, the Reynolds and Nusselt numbers are predicted to scale like $\re\sim \sqrt{\ra}$ and $\Nu\sim \sqrt{\ra}$, with different $\Pra$ scaling depending on the theory~(\cite{kra62,spi71,gro00,gro01,gro02}).  The question of the existence of this regime was addressed in simulations of homogeneous Rayleigh-B\'{e}nard convection in a tri-periodic cell (i.e., {\it without confinement}) 
where evidence was found supporting such a regime when all of the boundaries (and thus boundary layers) were absent~(\cite{loh03,cal05,cal06}) and confirming (\cite{cal05}) the Prandtl number dependence as 
suggested by \cite{gro00,gro01,gro02}.  
  Exponentially growing solutions at low $\ra$ were discovered analytically
 by \cite{cal06} 
and observed in the simulations, appearing as ``elevator modes'' composed of strong upwards and downwards jets, and coinciding with an increasing heat transfer until the modes became unstable and disintegrated. 

Later, experiments performed in long rectangular channels by \cite{gib06,gib09,tis10}) and 
cylindrical pipes by \cite{ara09} also found scaling laws consistent with those expected for the ultimate regime, despite the presence of the side walls which cause an additional anisotropy and drag in the flow.  They also observed a mean global flow composed of hot upwards- and cool downwards-moving columns.  Whereas in standard RBC strong thermal gradients occur at the top and bottom plates, leaving the temperature field within the bulk flow nearly uniform, in the experiments with long cells, there exists a mean, linear temperature gradient throughout the bulk.  It is this underlying linear gradient which is used to drive the convection in homogeneous RBC, and in the axially-periodic convection cell considered here.

In this work, we numerically and analytically investigate thermal convection in a long vertical cylinder,  in both the low $\ra$ and high $\ra$ limits.  Though the top and bottom boundary layers are absent, the side walls are still present as in the experimental situations.  
The objective is to find out whether also in this confined (and presumably more stabilizing) geometry
exponentially growing solutions (so called elevator-modes) exist, how the Nusselt and Reynolds number depend on $\ra$, and 
how the results compare to recent experiments in confined flow geometries 
(\cite{per02,gib06,gib09,ara09}).
We first derive  exponentially growing mode solutions of the governing Boussinesq equations with 
laterally confined boundary conditions, and compare them with results from simulations. Then we increase $\ra$ and  find that interaction between the modes stabilizes the system. This interaction
 prevents the growth of the exponential modes, and allows us to define 
the global heat transfer and the global Reynolds number of the system.

\section{The axially-homogeneous Rayleigh-B\'{e}nard system and a class of exact analytic solutions}

\subsection{Definition of the system}
The  system under study is axially-homogeneous thermal convection in a vertical cylinder, i.e.,  in a domain with periodic 
boundary conditions in the axial direction,  but with lateral confinement.
Standard Rayleigh-B\'{e}nard convection occurs in a cylinder (or other geometry) with a top and bottom plate kept at fixed temperatures, and this temperature difference drives the flow.  In the homogeneous or periodic case, with no boundaries except the radial wall, we drive the flow using a temperature gradient $\Delta'=\rd\bar{T}/\rd z$, where $\bar{T}$ is an underlying linear temperature profile in $z$ about which fluctuations $\Theta$ occur.  The absolute local temperature can then be written 
as $T(r,\theta,z,t)=\bar{T}(z)+\Theta(r,\theta,z,t)$. This linear temperature (or density gradient in the case of~\cite{ara09}) is observed in 
experiments~(\cite{gib06,gib09}). The diameter of the cylinder $d$ is the only imposed scale, and therefore, we make the hypothesis that $d$ is the relevant length scale for the convection (aside from the additional smaller scales as turbulence develops).
The periodicity length $L$,  which can be combined with the diameter to form the aspect ratio  $\Gamma = d/L$, is much less relevant because it does not change the forcing of the system.
We non-dimensionalize the governing Navier-Stokes equations using $\hat{x}=d$ for lengths, the free-fall speed $\hat{u}=\sqrt{\alpha g \Delta' \, d^2}$ for velocities, and $\hat{T}=\Delta' \, d$ for temperatures.  The Prandtl number is $\Pra=\nu/\kappa$ and the Rayleigh number is 
\beq
\ra=\frac{\alpha g \Delta' \, d^4}{\nu \kappa}.
\eeq
Here the material parameters are  the coefficient of thermal expansion $\alpha$, the kinematic viscosity  $\nu$, and the thermal diffusivity $\kappa$;   the gravitational acceleration is denoted as $g$. Notice that the same definition of the Rayleigh number was chosen in the experimental studies by \cite{gib06,gib09} and \cite{tis10}. The more usual Rayleigh number based on the height of a closed Rayleigh-B\'enard cell, $Ra_{L} = \alpha g \Delta T \ L^3 (\nu \kappa)^{-1}$, where $ \Delta T$ is the temperature difference
 between the horizontal plates, is linked to the present one through the relation $Ra_L = Ra\ \Gamma^4$.

In non-dimensional form, the Boussinesq equations read 
\bqa
\frac{\rD \bar{u}}{\rD t}&=&-\bar{\nabla} P+\sqrt{\frac{\Pra}{\ra}}\nabla^2\bar{u}+\Theta\, \hat{z} \nonumber \\ 
\frac{\rD \Theta}{\rD t}&=&\frac{1}{\sqrt{\Pra\, \ra}}\nabla^2\Theta+w
\eqa
where $\bar{u}=(u,v,w)$ is the velocity field with components in the $r$, $\theta$, and $z$ directions, respectively, and $P$ is the dimensionless pressure field.

The velocity field satisfies the no-slip condition at the lateral walls located (in non-dimensional units) at $r=r_{ext} =  1/2$ so that 
\beq
\bar{u}|_{r=1/2}=0.   \label{bcvel}
\eeq
For the temperature $T$ we consider either a fixed condition at the lateral wall, which in term of the fluctuations $\Theta$ means  
\beq
\Theta|_{r=1/2}=0,  \label{bctempfix}
\eeq
or adiabatic conditions at the wall, corresponding to:
\beq
\partial \Theta / \partial r |_{r=1/2}=0.   \label{bctemp}
\eeq
The system defined so far models the physical situation of a vertical pipe connected to two infinite reservoirs of fluids at different temperatures. However, in real experiments homogeneous convective systems are produced by connecting  a vertical conduct to two closed and finite chambers filled with fluid at different temperatures (see e.g. ~\cite{gib06} and \cite{ara09}).
Therefore, in order to mimic such a condition in the following it will be important to take into account the extra boundary condition of no mean vertical momentum and no mean temperature increase in the system 
\beq
\int_0^{r_{ext}} \int_{0}^{2 \pi} w \ dr\  d\theta = 0 ; \quad  \int_0^{r_{ext}} \int_{0}^{2 \pi} \Theta \ dr\  d\theta = 0; \label{nomeanflow}
\eeq
which shall be satisfied at any point $z$ on the vertical axis of the cylinder.
Note however that this ``no mean value''  condition does not prevent from having a positive heat flow, and indeed for the convective regime - as opposed to the conductive one -  we expect 
$\left< \int_0^{r_{ext}} \int_{0}^{2 \pi} w \Theta \ dr\  d\theta \right>_t > 0$.

\subsection{Exact solutions to governing equations}
As in the tri-periodic case, we look for exact solutions which are translationally invariant in $z$ and have $u=v=0$ ~(\cite{cal06}). The coupled equations for the vertical velocity $w(r,\theta,t)$ and temperature $\Theta(r,\theta,t)$ are then:
\bqa
\frac{\partial w}{\partial t}&=&\sqrt{\frac{\Pra}{\ra}}\nabla^2 w + \Theta, \nonumber \\ 
\frac{\partial \Theta}{\partial t}&=&\frac{1}{\sqrt{\Pra\, \ra}}\nabla^2 \Theta + w.
\label{gov}
\eqa
An adiabatic side wall boundary condition (Eq.~\ref{bctemp}) is relevant for experiments where tanks or cells are insulated at the sides, and will be implemented throughout the remainder of this paper.  Appendix~\ref{app} gives the analytic result for the fixed temperature side wall boundary condition (Eq.~\ref{bctempfix}).  The mixed (\textit{i.e.} adiabatic) boundary condition case is more difficult to solve analytically, but solutions can still be written for the special case $\Pra=1$. This case will be described in detail in the following.

 Trying solutions with an exponential growth time dependence ($e^{\lm t}$), Eq.~\ref{gov} simplifies and can be uncoupled by introducing two new fields $X(r,\th)$ and $Y(r,\th)$
\bqa
X(r,\th)&=&w(r,\th)-\Theta(r,\th) \nonumber \\
Y(r,\th)&=&w(r,\th)+\Theta(r,\th) 
\eqa
which satisfy
\bqa
\sqrt{\ra} (\lambda+1)X&=&\frac{\p^2 X}{\p r^2}+\frac{1}{r}\frac{\p X}{\p r}+\frac{1}{r^2}\frac{\p^2 X}{\p \theta^2}, \nonumber \\ 
\sqrt{\ra} (\lambda-1)Y&=&\frac{\p^2 Y}{\p r^2}+\frac{1}{r}\frac{\p Y}{\p r}+\frac{1}{r^2}\frac{\p^2 X}{\p \theta^2}. \label{newgov}
\eqa
The azimuthal dependence can be expanded as a combination of $\sin(n\th)$ and $\cos(n\th)$ and written in terms of the modes $X_n$ and $Y_n$ which satisfy the modified Bessel equation:
\bqa
r^2\frac{\rd^2 X_n}{\rd r^2}+r\frac{\rd X_n}{\rd r}-\left(\sqrt{\ra}(\lmn+1)\, r^2+n^2 \right)X_n&=&0, \nonumber \\
r^2\frac{\rd^2 Y_n}{\rd r^2}+r\frac{\rd Y_n}{\rd r}-\left(\sqrt{\ra}(\lmn-1)\, r^2+n^2 \right)Y_n&=&0. \label{bess}
\eqa
The $m$ label for the growth rate is necessary because again there are multiple solutions for a given $n$, $\lmn$, which satisfy the boundary conditions. The radial dependencies of the solutions to Eq.~\ref{bess} are
\bqa
X_n(r)&=&x_0\, I_n\left(\sqrt{(\lmn+1)}\,\ra^{1/4}\, r \right) \nonumber \\
Y_n(r)&=&y_0\, I_n\left(\sqrt{(\lmn-1)}\,\ra^{1/4}\, r \right),  \label{spatial}
\eqa
where $I_n$ is the modified Bessel function of the first kind.  The modified Bessel functions of the second kind, $K_n$, are also solutions to Eq.~\ref{newgov} but are not physically acceptable because they diverge at $r=0$.

The vertical velocity and temperature fields then have the form
\bqa
w_n(r,t)&=& a_0\ e^{\lmn t}\ [ Y_n(r)+X_n(r) ] \nonumber \\
\Theta_n(r,t)&=& a_0\ e^{\lmn t}\ [ Y_n(r)-X_n(r) ],
\label{solform}
\eqa
where the prefactor $a_0$ is unprescribed. 

The growth rate of each mode $\lmn$ depends on $\ra$. After applying the boundary conditions at the cylinder wall at $r=r_{ext}=1/2$ (Eqs.~\ref{bcvel} and~\ref{bctemp}), the solutions correspond to the roots of 
\beq
-\sqrt{\frac{\lm-1}{\lm+1}}\cdot \frac{I_n (\frac{1}{2}\sqrt{(\lm+1)}\,\ra^{1/4})}{I_n (\frac{1}{2}\sqrt{(\lm-1)}\,\ra^{1/4})}=\frac{I_{n-1}(\frac{1}{2}\sqrt{(\lm+1)}\,\ra^{1/4})+I_{n+1}(\frac{1}{2}\sqrt{(\lm+1)}\,\ra^{1/4})}{I_{n-1}(\frac{1}{2}\sqrt{(\lm-1)}\,\ra^{1/4})+I_{n+1}(\frac{1}{2}\sqrt{(\lm-1)}\,\ra^{1/4})}. \label{lambdamn}
\eeq
Even though there are multiple solutions $\lm=\lmn$ for each mode $n$, the maximum growth rate will dominate over the others as time progresses. 

Intuitively one might guess that the exponential growth cannot be maintained, and would be arrested by friction on the side walls.  However, the solutions to the above system of equations have not ignored viscous drag on the walls.  Even though the velocity, and gradients at the walls, are growing exponentially, the frictional force does not overtake the driving buoyancy force. 

In the periodic RB cell, the global, time-averaged dimensionless heat transfer is given as 
\beq
\Nu=\frac{\left<w\ \Theta \right>}{\kappa\ \Delta'}+1,
\label{nuseq}
\eeq
where the brackets indicate a volume (surface would be the same) and time average.  When $w\sim e^{\lmn t}$ and $\Theta\sim e^{\lmn t}$, the instantaneous heat transfer $\Nu(t)\sim e^{2 \lmn t}$ also grows exponentially in time.

\begin{table}
\begin{center}
\begin{tabular}{@{\extracolsep{8pt}}c|c|ccc}
\hline
$\ra$ &$\lambda_{mn}$ & $m=1$ & $2$ & $3$ \\
\hline
$7.66\cdot 10^3$ &$n=0$     &  0.81  &  0.022 & --- \\
			   & 1     &  0.52  &  --- & --- \\
			   & 2     &  0.14  &  --- & --- \\
\hline
$3.06\cdot 10^4$ & 0     &  0.89  &  0.45 & --- \\
& 1     &  0.73  &  0.12 & --- \\
& 2     &  0.52  &  --- & --- \\
& 3     &  0.26  &  --- & --- \\
\hline
$1.23\cdot 10^5$ & 0     &  0.94  &  0.70 & 0.28 \\
& 1     &  0.86  &  0.52 & 0.0041 \\
& 2     &  0.74  &  0.31 & --- \\
& 3     &  0.60  &  0.080 & --- \\
& 4     &  0.44  &  --- & --- \\
& 5     &  0.25  &  --- & --- \\
& 6     &  0.040  &  --- & --- \\
\hline
\end{tabular}
\caption{Positive growth rates $\lmn$ corresponding to active modes, 
for $\ra=7.66\cdot 10^3$ (top), $\ra=3.06\cdot 10^4$ (middle) and $\ra=1.23\cdot 10^5$ (bottom) at $\Pra=1$.}
\label{growthrates}
\end{center}
\end{table}

\subsubsection{Axisymmetric, dipolar modes and critical Rayleigh number}
The spatial form of solutions corresponding to the zero mode case ($n=0$) have an axisymmetric profile (see Eq.~(\ref{spatial})). 
This means that an increasing amount of fluid is transported through the cylinder and that the mean temperature of the system increases in time. Obviously these zero-mode axisymmetric solutions have a non-zero mean vertical flow and non-zero mean temperature fluctuation,  therefore they are in conflict with the condition (\ref{nomeanflow}).
Although these solutions are non-physical (they correspond to an unbounded infinite system) they will be useful to validate our simulations.

From the above considerations we deduce that the first unstable physical mode is $n=1$, which corresponds to a dipolar flow profile, moving upwards in half of the cylinder and downwards in the other half.  
Such dipolar modes were also observed 
in experiments in a cylindrical column~(\cite{ara09}).
The critical Rayleigh number corresponding to $\lambda_{11}(\ra_c)=0$ can be computed from Eq. (\ref{lambdamn}), which leads to the value  $\ra_c = 1087.397$.
In Table~\ref{growthrates} we list the active modes at different $Ra$ numbers above $\ra_c$. For the specific case $\ra=7.66\cdot 10^3$ and $\Pra=1.0$, the $n=0$ maximum growth rate is $\lm_{10}=0.81$. The first physical dipolar mode has growth rate $\lm_{11}=0.52$. 
The growth rates and their spatial dependencies are confirmed in the following simulations.


\section{Numerical Method}
Numerical simulations in a cylindrical cell are performed using the code developed by \cite{ver96} and \cite{ver03}, modified to satisfy the periodic conditions at $z=0,\ L$. Furthermore, to satisfy the condition (\ref{nomeanflow}) the vertical flow and mean temperature fluctuation are subtracted at each time step. A similar procedure was adopted in the earlier tri-periodic cell simulations by \cite{loh03,cal05,cal06}.
A finite difference scheme on a staggered grid is used to directly solve the incompressible continuity, momentum equation, and energy equations under the Boussinesq approximation using a fractional step method~(\cite{ver96,ver03,kim85}).  The grid is non-uniform in the radial direction, but is uniform in $z$ due to the periodicity, allowing the pressure solver to take advantage of trigonometric expansions and making the routine more efficient. 
In the simulations we must fix the axial periodicity length $L$, which is here just a numerical parameter rather than a physical scale for the system dynamics.  We expect, and will find, that as the aspect ratio $\Gamma=d/L$ is reduced (by increasing the periodicity length $L$), the system properties become independent of $\Gamma$.

All the simulations have been started with a velocity field at rest and the temperature field consisting only of a random perturbation 
of maximum amplitude $0.01 \Delta'd$.

\begin{figure}
\begin{center}
\psfrag{wmax}{\large{$w_{max}$}}
\psfrag{t}{\large{$t$}}
\psfrag{r}{\large{$r$}}
\psfrag{tw}{\large{$w(r),\ \Theta (r)$}}
\psfrag{wsimulation}[r][r]{\footnotesize{$w-$simulation}}
\psfrag{thsimulation}[r][r]{\footnotesize{$\Theta-$simulation}}
\psfrag{wtheory}[r][r]{\footnotesize{$w-$theory}}
\psfrag{ththeory}[r][r]{\footnotesize{$\Theta-$theory}}
	\includegraphics[width=0.48\textwidth, angle=0,trim=0.38in 0 0 0,clip]{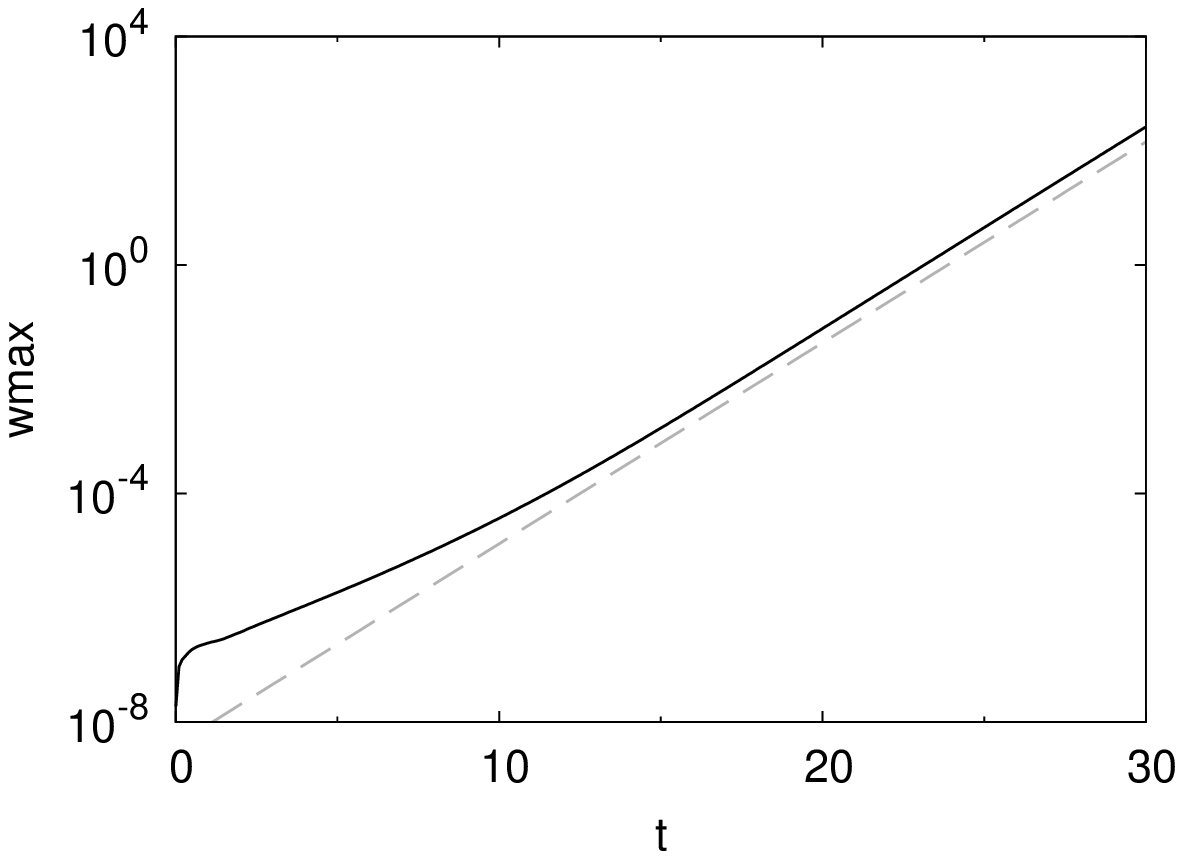}
$\quad$
	\includegraphics[width=0.48\textwidth, angle=0,trim=0.45in 0 0 0,clip]{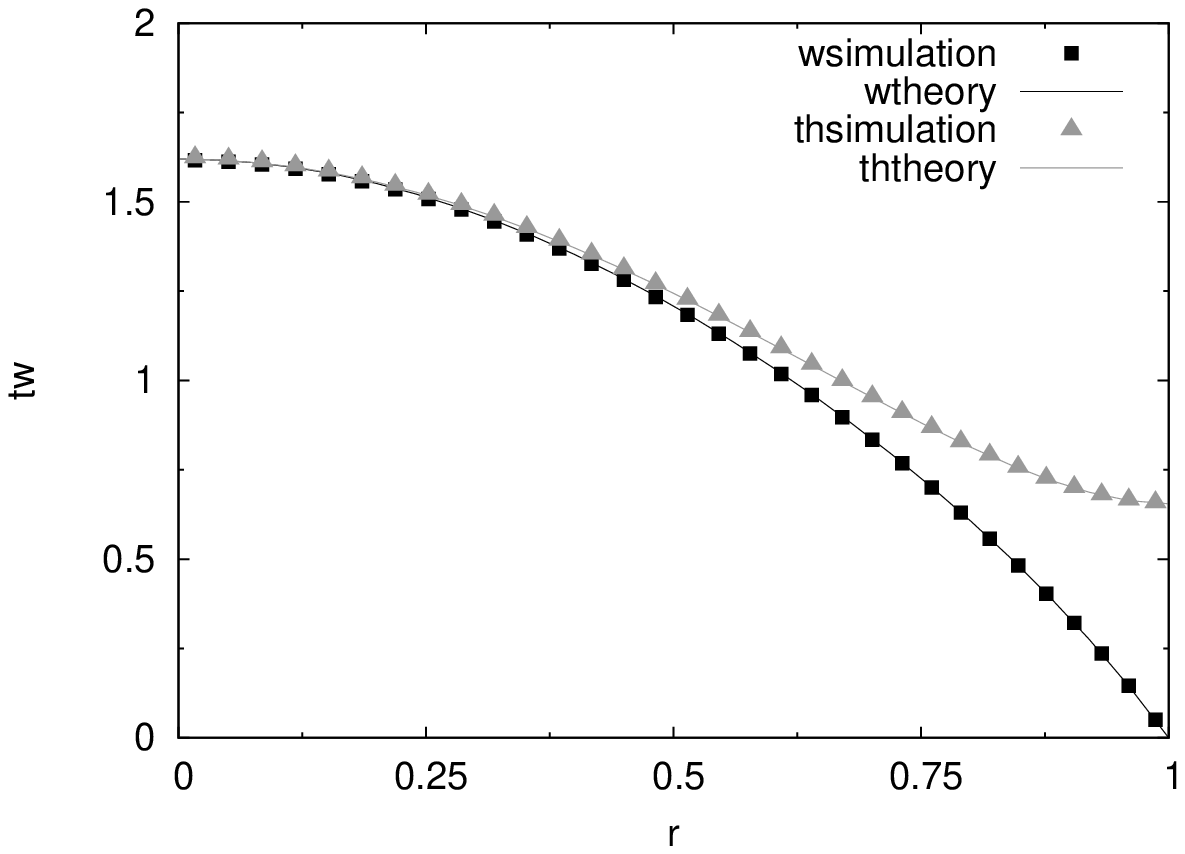}
\caption{Growth rate (a) of vertical velocity field $w$ for $n=0$ axisymmetric mode as function of non-dimensional time -- simulation with $\Gamma=1/2$ (solid black line) approaches theoretical prediction of  $\lambda_{10}=0.81$ (dashed grey line).  Spatial forms (b) of $w$ and temperature fluctuation field $\Theta$ for $\ra=7.66\cdot 10^3$, simulation and theory coincide. }
\label{aximode}
\end{center}
\end{figure}

\begin{figure}
\begin{center}
\psfrag{wmax}{\large{$w_{max}$}}
\psfrag{t}[c][t]{\large{$t$}}
\psfrag{r}{\large{$r$}}
\psfrag{tw}{\large{$w(r),\ \Theta (r)$}}
\psfrag{wsimulation}[r][r]{\footnotesize{$w-$sim.}}
\psfrag{thsimulation}[r][r]{\footnotesize{$\Theta-$sim.}}
\psfrag{wtheory}[r][r]{\footnotesize{$w-$theo.}}
\psfrag{ththeory}[r][r]{\footnotesize{$\Theta-$theo.}}
	\includegraphics[width=0.48\textwidth, angle=0,trim=0.38in 0 0 0,clip]{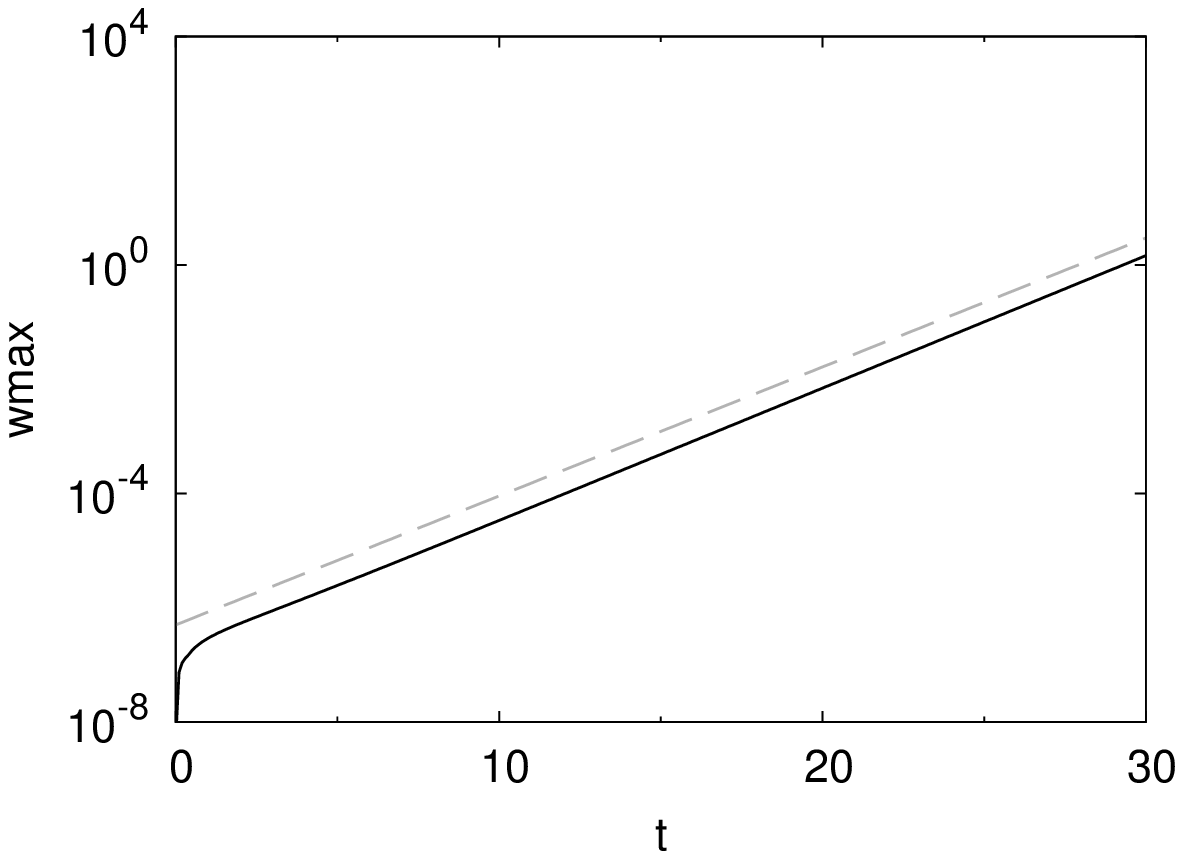}
$\quad$
	\includegraphics[width=0.48\textwidth, angle=0,trim=0.45in 0 0 0,clip]{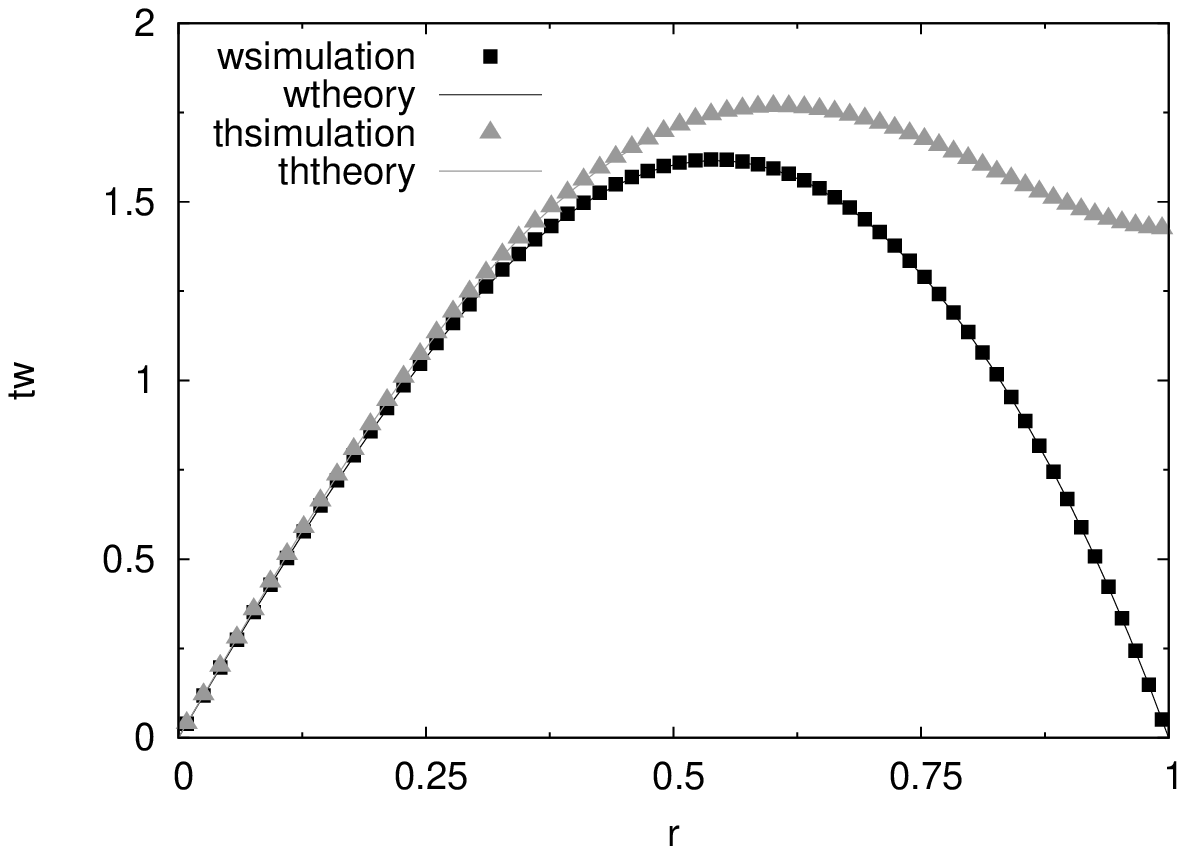}
\caption{Growth rate (a) of vertical velocity field $w$ for $n=1$ dipolar mode as function of non-dimensional time -- simulation with $\Gamma=1/2$ (solid black line) approaches theoretical prediction of $\lambda_{10}=0.52$ (dashed grey line).  Spatial forms (b) of $w$ and temperature fluctuation field $\Theta$ for $\ra=7.66\cdot 10^3$, simulation and theory coincide.}
\label{dimode}
\end{center}
\end{figure}

The new code implementation with periodic conditions was first checked by direct comparison of the predicted growth rates and radial forms of $w$ and $\Theta$ from the analysis in the previous section, both for cases where $n=0$ and $n=1$ modes dominated, along with the fixed temperature side wall boundary conditions (Appendix~\ref{app}).  For $\ra=7656$, $\Pra=1$,  and $\Gamma=1/2$, the axisymmetric and dipolar modes are isolated at early times and are shown in Figs.~\ref{aximode} and~\ref{dimode}, comparing well to the theoretical predictions.  Here, because the velocity and temperature fields are observed to be $z$-independent, we use the maximum vertical velocity field calculated at each time step to represent $a_0\ e^{\lmn t}$ from Eq.~\ref{solform}. 

\begin{figure}
\begin{center}
\psfrag{wmax}{\large{$w_{max}$}}
\psfrag{t}[c][t]{\large{$t$}}
\includegraphics[width=0.68\textwidth, angle=0,trim=0.3in 0.9in 0 0.5in,clip]{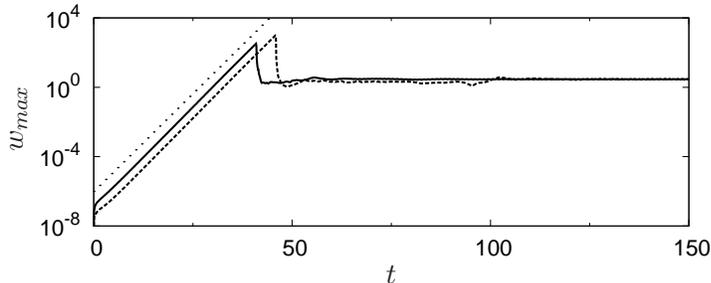}
\caption{Exponential growth and stabilization of vertical velocity field $w$ for $\ra=7.66\cdot 10^3$, $\Pra=1$, and $\Gamma=1/4$ as function of non-dimensional time, for resolutions $65\times33\times129$ (solid line) and $129\times65\times129$  (dashed line). Predicted growth rate of $\lm_{11}=0.52$ is shown as dotted line.}
\label{rescheck}
\end{center}
\end{figure}

When only a few of the exponentially growing modes are active at low $\ra$ (as for the parameters in Table~\ref{growthrates}), the velocity and temperature gradients at $\th_n=2\pi/n$ grow exponentially and thus the resolution unavoidably becomes insufficient at some point.  However, this is only an issue for low $\ra$ cells because as more modes are available, their interaction prevents such a situation from occurring.  We find that though the resolution can affect the maximum $w$ and $\Theta$ (and thus, the heat transfer) obtained during the oscillations of the system, the growth rates are unaffected.  For example, at $\ra=7656$, $\Pra=1$,  and $\Gamma=1/4$ where the $\lm_{11}$ mode strongly dominates the system, simulations with a resolution of $129\times65\times129$ (in $\theta$, $r$, $z$) reach a maximum vertical velocity of $w_{max}=1000$ before stabilizing, compared to $w_{max}=200$ for a resolution of $65\times33\times129$ (shown in Fig.~\ref{rescheck}).  Even though the early time dynamics depend on the resolution, the growth rate and steady-state that is reached are unaffected.
The situation is analogous to what \cite{cal06}
 had found in the unconfined RB system.

The consistency of the code was also verified using the relation between the Nusselt number and the volume and time-averaged kinetic dissipation rate, $\epsilon_u=\nu|\nabla \mathbf{u}|^2$.  For the periodic cell,  $\epsilon_u=\nu^3\ \Nu \ \ra \ \Pra^{-2}\ d^{-4}$ which is confirmed within few percents as long as the spatial resolution of the simulation is adequate (see Table~\ref{fulltab}). It should be noted that for increasing $Ra$ the previous identity is verified less satisfactorily since limitations of the computational resources prevented us from running cases on larger grids. 

\section{Results}
\begin{figure}
	\psfrag{z}{\large{$z$}}
	\psfrag{uz}{\large{$w$}}
        \psfrag{x}[c][c]{\large{$x$}}	
\centering
\subfigure[][\ $t=99.90$]{	
	\includegraphics[width=0.23\textwidth, angle=0,trim=4.1cm 1cm 4.05cm 1.0cm, clip=true]{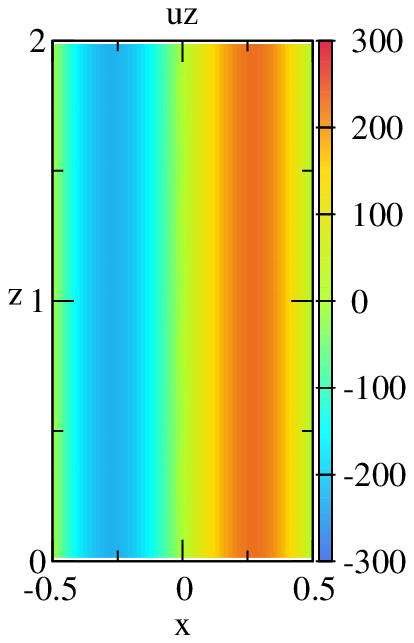}}
\subfigure[][\ $t=99.96$]{	
	\includegraphics[width=0.23\textwidth, angle=0,trim=4.1cm 1cm 4.05cm 1cm, clip=true]{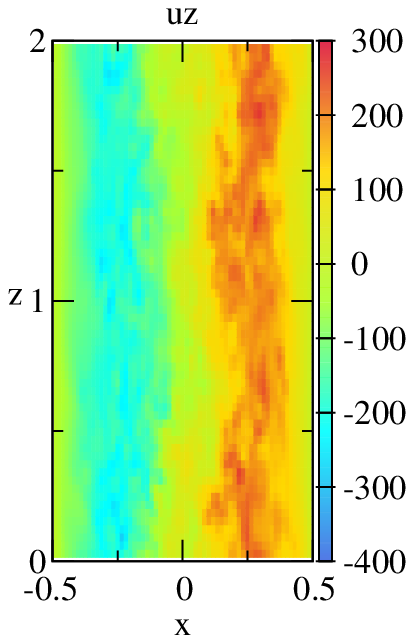}}
\subfigure[][\ $t=100.02$]{	
	\includegraphics[width=0.23\textwidth, angle=0,trim=4.1cm 1cm 4.05cm 1cm, clip=true]{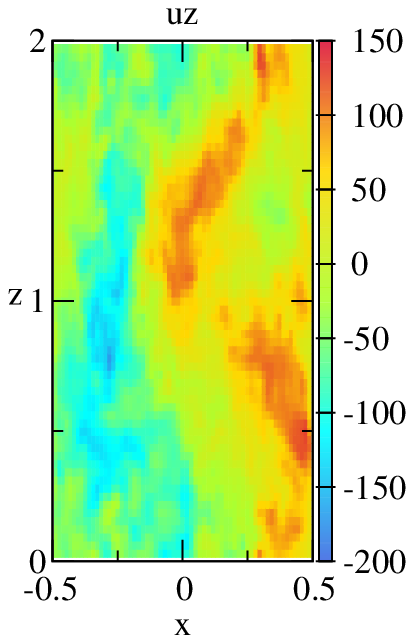}}
\subfigure[][\ $t=100.2$]{	
	\includegraphics[width=0.23\textwidth, angle=0,trim=4.1cm 1cm 4.05cm 1cm, clip=true]{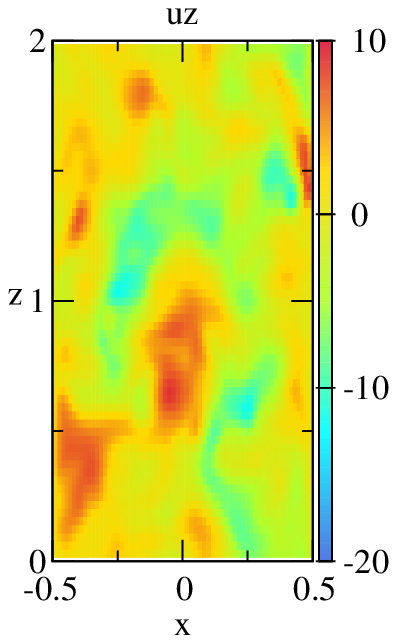}}
\caption{Instantaneous contour plots of the dimensionless vertical velocity $w$, through a vertical cross section of the cell, for $\ra=7.66\cdot 10^3$, $\Pra=1$, and $\Gamma=1/2$. Times are chosen just before stabilization of the growing dipolar mode at $t=99.9$ (a), during stabilization when horizontal velocity fluctuations are growing at $t=99.96$ (b) and $t=100.02$ (c), and later after decay at $t=100.2$ (d).  Note that the colorbar scales are varied according to the flow so that the structure can be seen at all times.  A movie of the process is provided in the Supplementary Material.}
\label{snapshots}
\end{figure}

\begin{figure}
\psfrag{uw}[c][c]{\large{$\uhrms,\ \wrms$}}
\psfrag{t}[c][t]{\large{$t$}}
\centering
\includegraphics[width=0.68\textwidth, angle=0,trim=0.3in 0.6in 0 0.4in,clip]{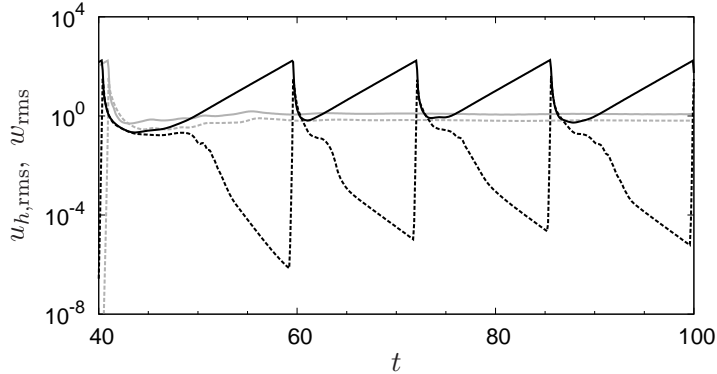}
\caption{Aspect ratio dependence of mode growth and decay at $\ra=7.66\cdot 10^3$, $\Pra=1$. Plotted are the r.m.s. vertical (solid line) and horizontal ($u_h$, dashed line) velocity components for $\Gamma=1/2$ (black) and $\Gamma=1/4$ (grey).}
\label{growdecay}
\end{figure}

At low $\ra$ we observe individual growing modes which, depending on $\ra$ and $\Gamma$, dominate the system dynamics.  Already shown in Figs.~\ref{aximode}--\ref{rescheck} is the exponential growth at early times for simulations with $\ra=7.66\cdot 10^3$, $\Pra=1$, $\Gamma=1/2$ and $1/4$.  After the initial growth at $\Gamma=1/2$, Fig.~\ref{growdecay} shows that the system temporarily stabilizes as horizontal velocity field fluctuations increase, which destroys the axially-uniform modes of Eq.~\ref{gov}.  This process continues periodically, with the horizontal field growing and decaying.  This is visualized in Fig.~\ref{snapshots} which shows snapshots of the dimensionless vertical velocity in a vertical cross section through the axis of the cell, just before, during, and after the stabilization process has occurred.  Analogous dynamics were observed in the tri-periodic homogeneous RB system of \cite{cal06}, and the process was found to depend sensitively on the numerical resolution and time-step size.  Here, we have an additional parameter, the periodicity length which can affect this stabilization process.  If the periodicity length is increased so that $\Gamma=1/4$, instead of observing the long-term periodic behavior, the system settles and stabilizes after the horizontal fluctuations become of the same order as the vertical fluctuations (Fig.~\ref{growdecay}).  This signals that the origin of the physical process which causes the destruction of the exponentially growing modes, is related to the ability of the fluctuations to extend along the $z$-direction. 

\subfigcaptopadj=-3pt
\subfigtopskip=8pt

\vspace{0.3in}
\subfigtopskip=-20pt

\begin{figure}
\psfrag{dlnnudt}[c][c]{\large{$\frac{1}{2}d\ln(\Nu)/dt$}}
\psfrag{t}[c][t]{\large{$t$}}
\centering
\includegraphics[width=0.68\textwidth, angle=0,trim=0.3in 0.3in 0 0.3in,clip]{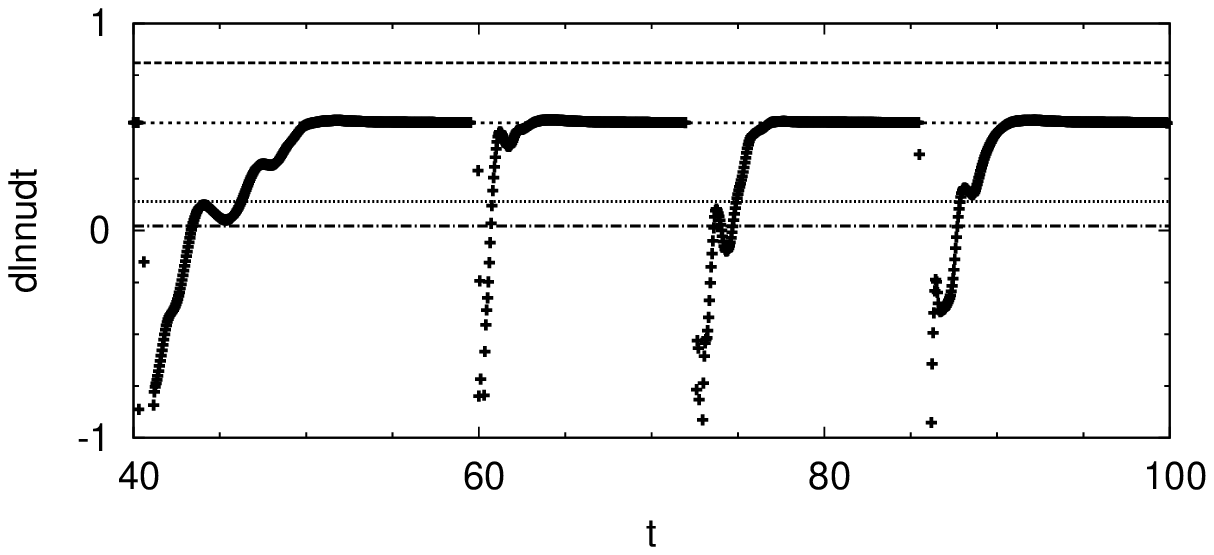}
\caption{Plot of $\frac{1}{2}d\ln(\Nu)/dt$, along with the active mode growth rates from Table~\ref{growthrates} for $\ra=7.66\cdot 10^3$ and $\Pra=1$, for $\Gamma=1/2$. The axisymmetric mode with $\lm_{10}=0.81$ was suppressed, so the system periodically sets in on the next fastest growing mode $\lm_{11}=0.52$.}
\label{dlnnudt7656}
\end{figure}

Plotting $\frac{1}{2}d \ln(\Nu)/dt$ as a function of time, along with the active $\lmn$, clearly shows convergence of the system to the fastest growing mode for $\ra=7.66\cdot 10^3$ with $\Gamma=1/2$ in Fig.~\ref{dlnnudt7656}.

\subfigcaptopadj=-23pt
\subfigtopskip=-20pt
\begin{figure}
\begin{center}
\psfrag{dlnnudt}[c][c]{\large{$\frac{1}{2}d\ln(\Nu)/dt$}}
\psfrag{t}[c][t]{\large{$t$}}
	\includegraphics[width=0.65\textwidth, angle=0,trim=0.3in 0.3in 0 0.5in,clip]{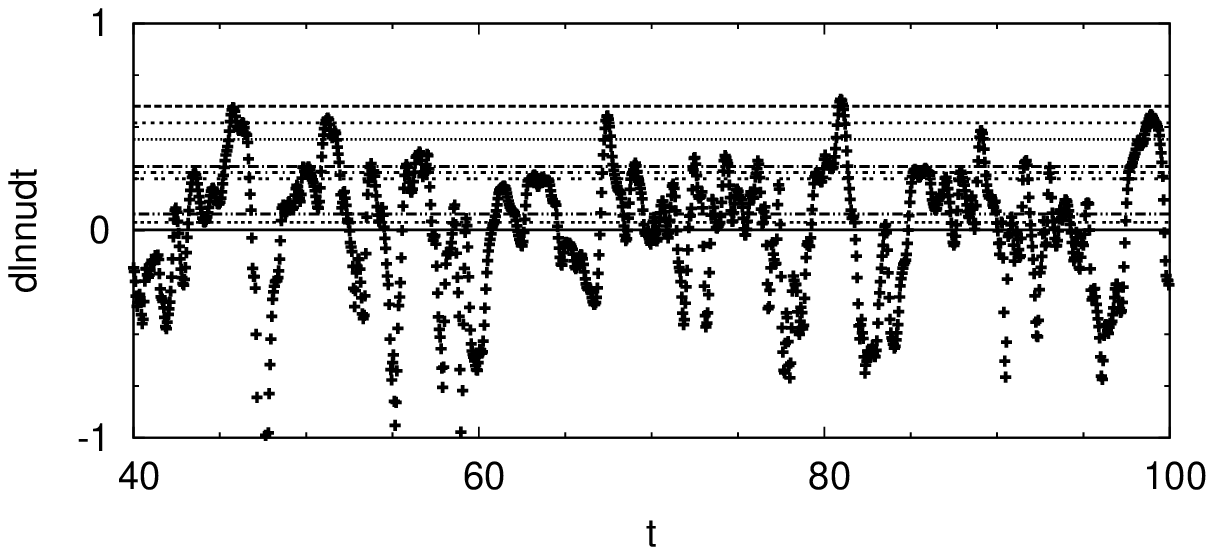}
	\includegraphics[width=0.65\textwidth, angle=0,trim=0.3in 0.3in 0 0.5in,clip]{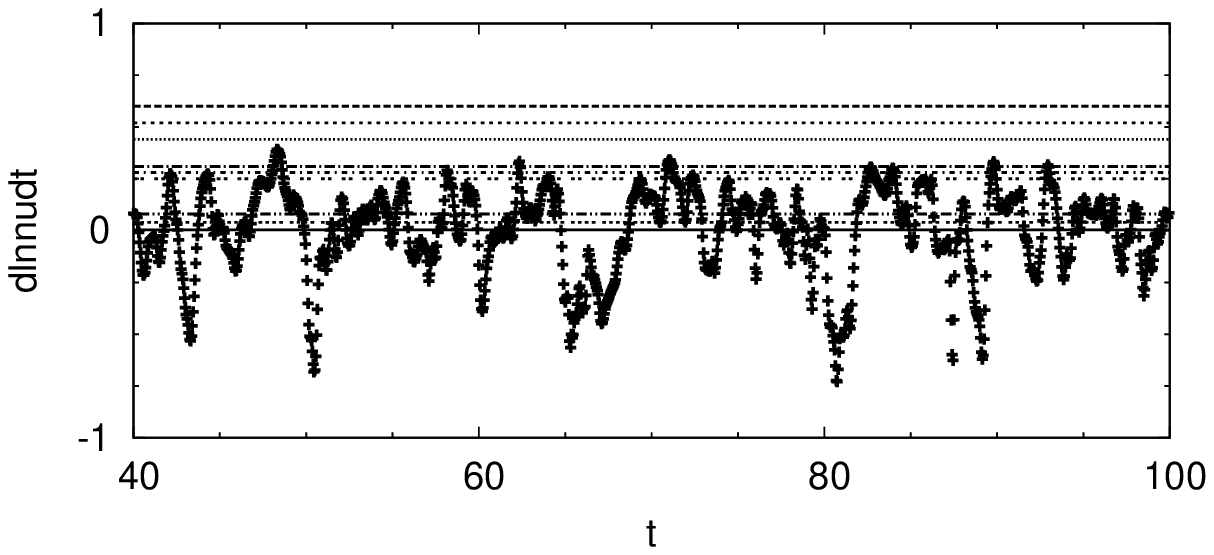}
\caption{
Plot of $\frac{1}{2} d \ln(\Nu)/dt$ for $\ra = 1.23 \cdot10^5$ and $\Pra = 1$, for 
$\Gamma=1/2$ (panel a) and $\Gamma=1/4$ (panle b). The active mode growth rates reached by the system are shown by the horizontal lines at $\lambda = 0.60, 0.52,
0.44, 0.31, 0.28, 0.25, 0.080, 0.04, 0.0041$ from Table~\ref{growthrates}.
}
\label{dlnnudt122500}
\end{center}
\end{figure}

\begin{figure}
\begin{center}
\includegraphics[width=1.0\textwidth, angle=0,trim=0.0in 0.0in 0.0in 0.0in, clip]{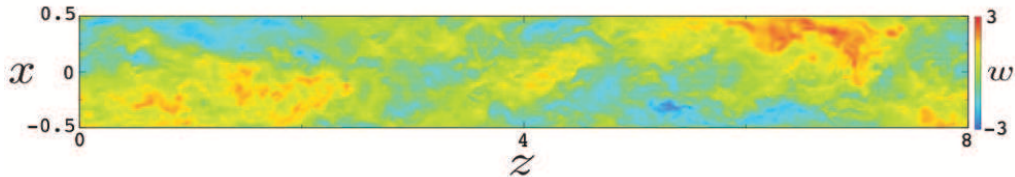}
\caption{Instantaneous contour plot of the dimensionless vertical velocity $w$, through a vertical cross section of the cell, for a typical state at $\ra=7.84\cdot 10^6$, $\Pra=1$, and $\Gamma=1/8$ }
\label{field0125}
\end{center}
\end{figure}

As $\ra$ is increased, more modes are present and the growth rates become closer together.  This leads to interaction between the modes and stabilization of the system by preventing individual modes from dominating the flow. Fig.~\ref{dlnnudt122500} shows the time variation of the Nusselt number for $\ra=1.23\cdot 10^5$ for $\Gamma=1/2$ and $\Gamma=1/4$.  Though the maximums jump among the lower growth-rate modes, there is no prolonged exponential growth as was seen at the lower $\ra$.  There is again a difference between $\Gamma=1/2$ and $1/4$ -- the lower aspect ratio simulations do not reach the same maximum growth rates, leading to a more stable system.  For reference and comparison to Fig.~\ref{snapshots} with $\Gamma=1/2$, Fig.~\ref{field0125} shows a snapshot of the vertical velocity field in a $\Gamma=1/8$ cell for $\ra=7.85\cdot 10^6$.

We have already seen that the aspect ratio of the cell, determined by the chosen numerical periodicity length, can affect the stabilization of the system by the breakdown of the exponentially growing modes (Figs.~\ref{growdecay} and~\ref{dlnnudt122500}).  Because we are interested in understanding the heat transfer and development of turbulence in a long pipe, when going to higher $\ra$ one would
 hope that the results are unaffected by the choice of the periodicity length.  Table~\ref{fulltab} shows simulation results at $\ra=O(10^{4}-10^{8})$ at aspect ratios $\Gamma=1/8$, $1/4$, and $1/2$.  The aspect ratio was changed by doubling the periodicity length, and keeping the resolution in $z$ the same. It appears that the global system properties indeed
start to converge once $\Gamma=1/4-1/8$ for $\ra=O(10^{5}-10^{6})$. For higher $\ra$, there is still aspect ratio dependence, but numerical limitations prevent us from increasing the periodicity length further. Since the $\Gamma=1/4$ simulations require half the number of grid points with respect to longer cells with $\Gamma=1/8$, higher $\ra$ simulations which already require a finer resolution are done with the intermediate aspect ratio $\Gamma=1/4$.

\begin{figure}
\begin{center}
\psfrag{urms}[c][c]{\large{$\uhrms,\ \wrms,\ \left< u_h w\right>$}}
\psfrag{r}[c][t]{\large{$x$}}
	\includegraphics[width=0.6\textwidth, angle=0,trim=0.473in 0.3in 0 0.5in,clip]{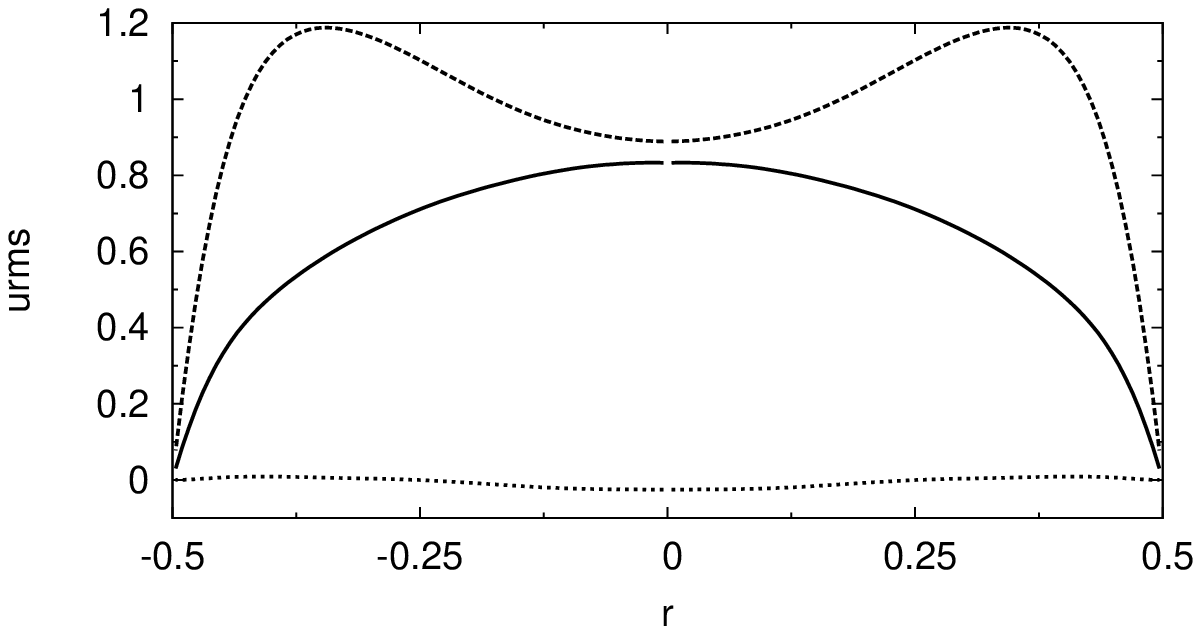}\\
	\includegraphics[width=0.6\textwidth, angle=0,trim=0.47in 0.3in 0 0.5in,clip]{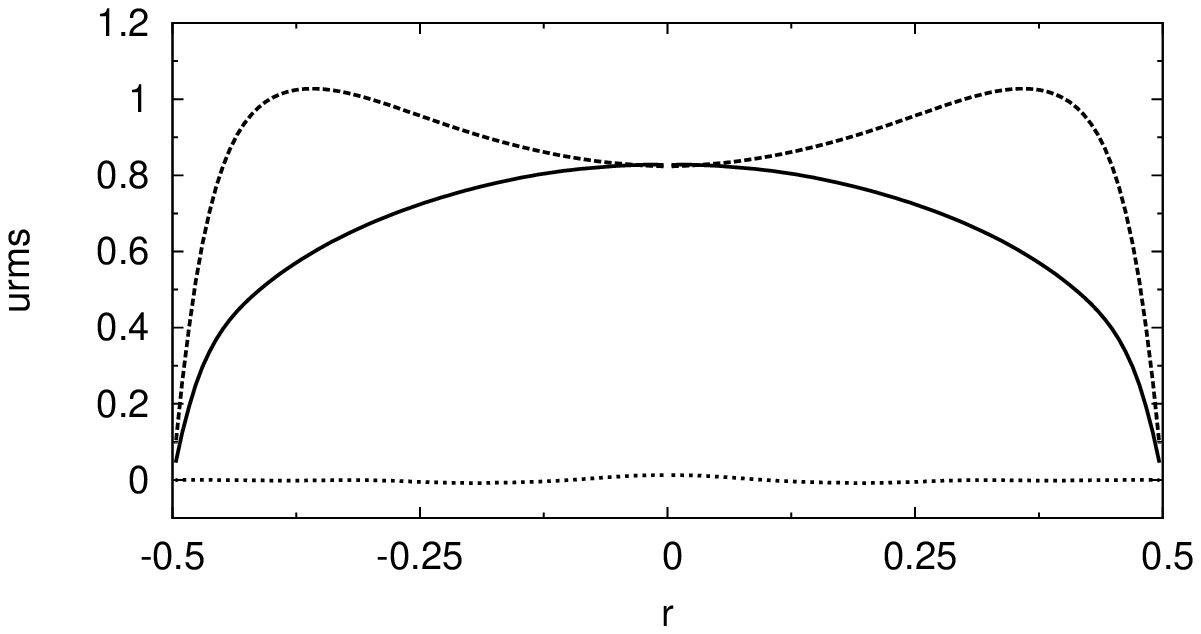}\\
	\includegraphics[width=0.6\textwidth, angle=0,trim=0.47in 0.3in 0 0.5in,clip]{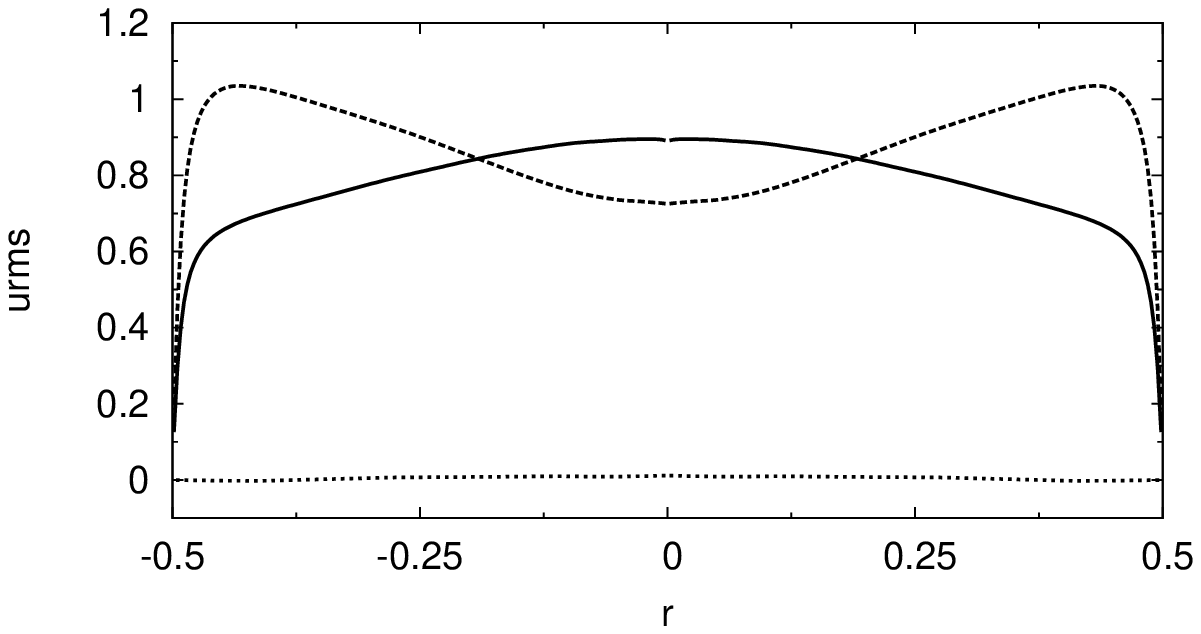}\\
	\end{center}
\caption{Spatial variation of r.m.s. horizontal ($\uhrms$, solid) and vertical ($\wrms$, dashed) velocites, and the turbulent shear stress ($\left< u_h w\right>$, dotted) for $\Gamma=1/4$ and $\Pra=1$, at $\ra=3.06\cdot 10^4$  (a), $\ra=1.23\cdot 10^5$  (b), and $\ra=1.57\cdot 10^7$ (c).}
\label{radprofs}
\end{figure}

The spatial variation of the r.m.s. velocity components plotted in Fig.~\ref{radprofs} for various $\ra$ shows the horizontal velocity $\uhrms$ peaks at the axis, and goes to zero at the walls. The vertical velocity fluctuations peak closer to the walls, with a local minimum along the axis, falling to zero at the walls owing to friction. 
At lower $\ra$, the horizontal fluctuations do not reach the same size as the vertical fluctuations at the axis. As $\ra$ is increased, $\uhrms$ increases relative to $\wrms$ at the axis, eventually overtaking $\wrms$.  The horizontal fluctuations also become more uniform across the cell.  The peaks in $\wrms$ become sharper and move towards the outside.  

Similar behavior was observed in the measured profiles in the buoyancy-driven pipe turbulence experiments of \cite{ara09}.  In their experiments with $\ra\approx10^8$ and $\Pra=670$, the profiles resembled those of Fig.~\ref{radprofs}(a) for the lower $\ra=3.06\cdot 10^4$ and $\Pra=1$, but the horizontal fluctuations reached only about 60\% the magnitude of the vertical fluctuations at the axis.  The reason for the resemblance to the lower $\ra$ result here is almost certainly the high value of $\Pra$. To check, we also calculated that the turbulent shear stress $\left<u_h w \right>$ was absent across the cell.  For all cases, $\left<u_h w \right>$  is close to zero compared to the r.m.s. components.  As in the experiments, the system remains anisotropic, evidenced  by the fact that $\left<w^2\right>/\left<u_h^2\right>\neq 1$, especially near the walls where the horizontal fluctuations fall to zero much more rapidly than the vertical fluctuations.
\subfigcaptopadj=-12pt
\subfigtopskip=-10pt

\begin{figure}
\psfrag{re}[c][t]{\large{$\re$}}
\psfrag{nu}[c][t]{\large{$\Nu$}}
\psfrag{ra}[c][t]{\large{$\ra$}}
\subfigure[][]{
	\includegraphics[width=0.5\textwidth, angle=0,trim=0.35in 0.2in 0 0.1in,clip]{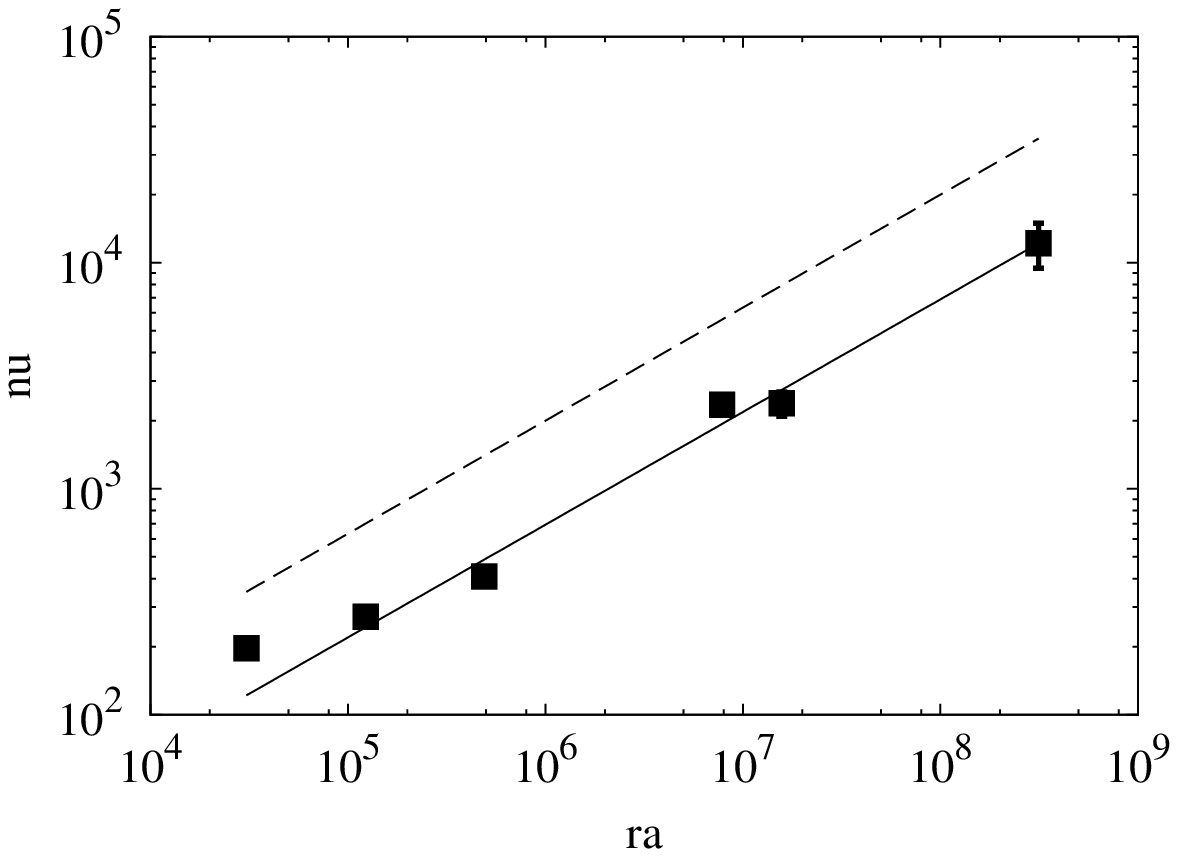}}
\subfigure[][]{
	\includegraphics[width=0.5\textwidth, angle=0,trim=0.35in 0.2in 0 0.1in,clip]{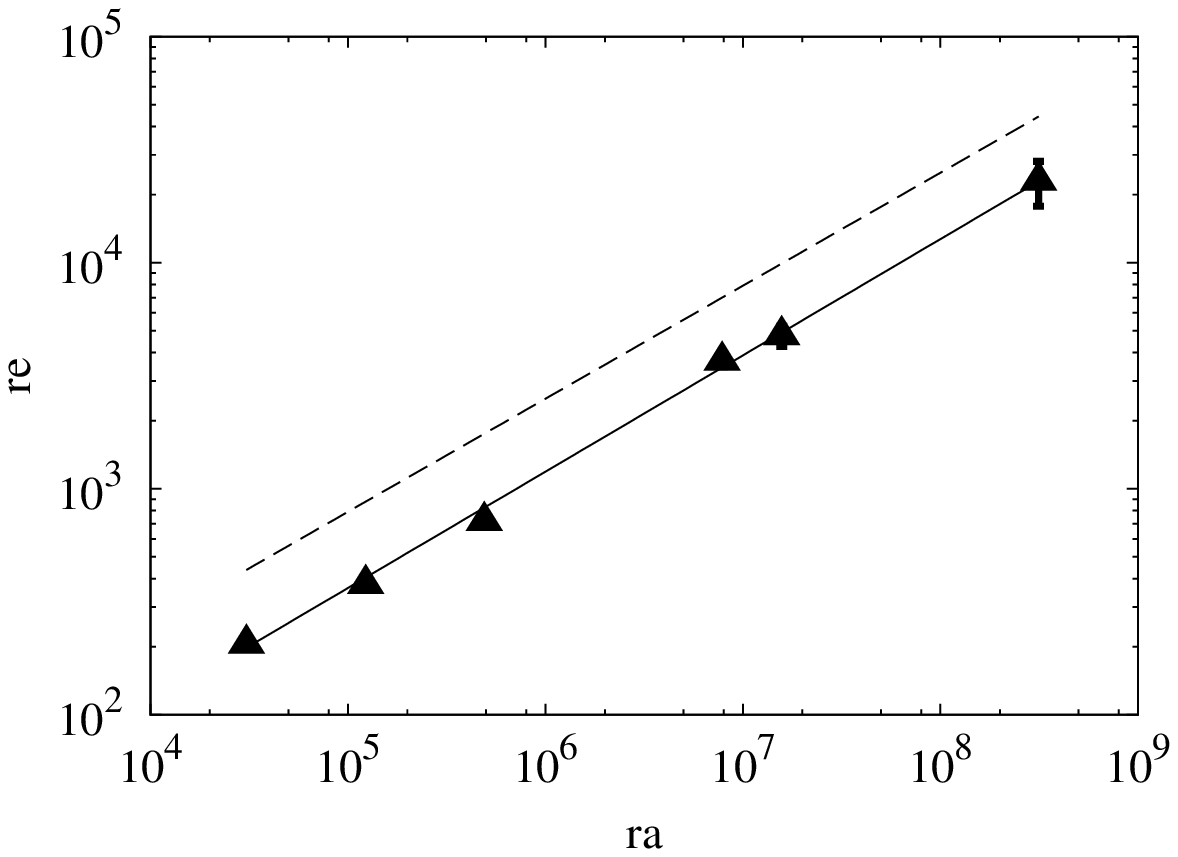}}
\caption{(a) Scaling of Nusselt number $\Nu$ and (b) Reynolds number $\re$ with $\ra$ for $\Gamma=1/4$ data. Dashed lines have slope $1/2$ for comparison. The power law fits (solid lines) have slopes of $0.50\pm 0.06$ for $\Nu$ and $0.51\pm 0.05$ for $\re$. }
\label{renu}
\end{figure}

Counterintuitively, the heat transfer at low $\ra=O(10^4)$ and smaller, 
is very large due to single exponentially modes dominating the system in this regime. For these systems, it is inappropriate to define an average $\Nu$ or $\re$ because we have seen that the maximum peaks (as in Fig.~\ref{growdecay}) depend on the grid resolution.   As $\ra$ is increased, the system stabilizes due to mode interaction and the scaling appears consistent with $\Nu\sim \ra ^{1/2}$ for $\ra=O(10^5-10^8)$, as shown in Fig.~\ref{renu}. The values used to determine the scaling law come from the aspect ratio $1/4$ runs (given in Table~\ref{fulltab}). Errors are estimated by the percentage difference of the two methods to calculate the Nusselt number (one, being the volume average in Eq.~\ref{nuseq}, the other being via the global relation with the kinetic energy dissipation rate). The highest $\ra$ runs appear under-resolved but nevertheless the prediction confirms the scaling laws.  The mean Reynolds number is calculated as $\re=(\left< u^2\right>_{V,t})^{1/2}\, d/\nu$ (where $u$ is the total dimensional velocity), and also appears consistent with  the $1/2$ power law from Fig.~\ref{renu}. 

\subfigcaptopadj=-12pt
\subfigtopskip=10pt

\begin{table}
\begin{center}
\begin{tabular}{@{\extracolsep{6pt}}cccccccc}
\hline
$\ra$ & $\Gamma$ & $N_{\theta}\ti N_{r}\ti N_z$ & $\Nu$ & $\re$ & $\left<w^2\right>/\left<u_h^2\right>$ & $\frac{\epsilon_{u}(\nabla_i u_j)}{\epsilon_{u}(Nu)}$ & $\eta/d$ \\
\hline
$3.06\cdot 10^4$ & 1/4 & $129\ti 65\ti 257$ & 197 & 205 & 3.12& 1.02& $2.06\cdot 10^{-2}$\\

$1.23\cdot 10^5$ & 1/2 & $129\ti 65\ti 129$ & 384 & 428 & 2.50 & 0.97 & $1.29\cdot 10^{-2}$	\\
$1.23\cdot 10^5$ & 1/4 & $129\ti 65\ti 257$ & 271 & 377 & 2.23 & 0.97 & $1.36\cdot 10^{-2}$	\\

$4.90\cdot 10^5$ & 1/2 & $257\ti 129\ti 257$ & 678	& 880 & -- & 1.00 & $7.95\cdot 10^{-3}$	\\
$4.90\cdot 10^5$ & 1/4 & $257\ti 129\ti 513$ & 409	& 716 & 1.84 & 1.00 & $8.54\cdot 10^{-3}$ \\
$4.90\cdot 10^5$ & 1/8 & $257\ti 129\ti 1025$ & 431 & 730 & 1.98 & 1.00 & $7.46\cdot 10^{-3}$\\

$7.84\cdot 10^6$ & 1/2 & $257\ti 129\ti 257$ & 4544 & 4702 & 2.49 & 0.832 & $2.79\cdot 10^{-3}$	\\   
$7.84\cdot 10^6$ & 1/4 & $257\ti 129\ti 513$ & 2349 & 3665 & 1.80 & 0.870 & $2.87\cdot 10^{-3}$\\
$7.84\cdot 10^6$ & 1/8 & $257\ti 129\ti 1025$	& 2196 & 3664 & 1.70 & 0.890 & $2.85\cdot 10^{-3}$\\

$1.57\cdot 10^7$ & 1/4 & $257\ti 129\ti 513$ & 2386 & 4747 & 1.59 & 0.83 & $2.4\cdot 10^{-3}$	\\
$1.57\cdot 10^7$ & 1/8 & $257\ti 129\ti 1025$	& 3248 & 5349 & 1.76 & 0.82 & $2.2\cdot 10^{-3}$ \\
$3.14\cdot 10^8$ & 1/4 & $257\ti 129\ti 1025$	& 12191 & 22909 & 1.43 & 0.55 & $8.5\cdot 10^{-4}$\\
\hline
\end{tabular}
\caption{Parameters and results from simulations.  Columns contain Rayleigh number $\ra$, aspect ratio $\Gamma$, grid resolution (number of nodes in $\theta$, $r$, and $z$ directions), Nusselt number $\Nu$ calculated by volume average of Eq.~\ref{nuseq}, Reynolds number $\re$ calculated from mean fluctuating velocity, ratio of the vertical to horizontal energies in fluctuating fields, check of global relation of $\epsilon_u(\nabla_i u_j)$ calculated directly from velocity gradients compared to $\epsilon_u(Nu) \equiv (\nu^3 \ \Nu \ \ra \ \Pra^{-2} d^{-4})$, and finally the non-dimensional Kolmogorov scale $\eta$ calculated from the kinetic energy dissipation rate $\epsilon_u(\nabla_i u_j)$.}
\label{fulltab}
\end{center}
\end{table}

\section{Conclusions}
We have presented theoretical results and simulations of thermal convection in a  laterally confined cylindrical pipe in the limit of small aspect ratios. At low $\ra$ we find exponentially growing modes consisting of upwards and downwards-flowing columns, analogous to those found in homogeneous RBC with no boundaries and simulations in tri-periodic cells~(\cite{cal05,cal06}), along with experimental observations~(\cite{gib06,gib09,ara09}).  The breakdown of these modes is found to depend on the numerical aspect ratio of the cell (or periodicity length), implying that their destruction is related to the ability of disturbances in the axial direction to grow.  At higher $\ra$, the scaling of the heat transfer and turbulence are consistent with the predicted (\cite{gro00,gro01,gro02,loh03}) 
$\Nu\sim\sqrt{\ra}, \re\sim\sqrt{\ra}$ scaling of the ultimate, bulk-dominated regime of thermal convection, just as in the case of the tri-periodic boundary conditions (\cite{loh03,cal05}).

It is remarkable that the sidewall boundary conditions 
and  the resulting kinematic boundary layers
do {\it not} lead to different scaling behavior
with a less steep increase of $\Nu$ with $\ra$, as common for the lower $\ra$ regimes, say 
$\ra \le 10^{12}$,
in the unifying theory of \cite{gro00,gro01,gro02}. 
The reason must be that these kinetic boundary layers only form at the sidewalls, and not at
the top and bottom walls, which do not exist in our setup. 
We therefore neither  expect 
 a less steep increase of $\Nu$ with $\ra$
beyond the onset of turbulence in 
the lateral kinetic boundary layers, which is expected to occur beyond  a critical
shear Reynolds number
$Re^*_s \approx 0.5 \sqrt{Re} \approx 420$ (\cite{ll87,gro00,gro02,gro11}). We note that here 
we are still considerably below this transition, as even for our largest $\ra$ we only have
$Re_s \approx 75 \ll Re^*_s$. 
It would be worth-while to push the numerical simulations in this laterally confined 
geometry to $\ra$ numbers
beyond the onset of the shear instability, in order to confirm that also then
$\Nu \sim \re \sim \ra^{1/2}$, without any logarithmic corrections, which 
seem to be typical for 
ultimate RB flow (and analogous ultimate Taylor-Couette flow)
in {\it fully confined} (i.e., with boundaries and boundary layers 
also towards the top and bottom)
geometries (\cite{cha97,cha01,dub01,gro11,gil11}).

\appendix
\section{Zero temperature fluctuations at side wall solution}
\label{app}
Though this boundary condition is not currently relevant for the experiments, it has a simple solution which was used to check the numerical results.  Zero temperature fluctuations at the side wall means that physically, the absolute temperature follows the imposed linear profile at the walls.  Eq.~\ref{gov} is exactly satisfied by solutions of the form
\bqa
w_{mn}(r,\theta,t)&=&w_0\ e^{\lmn t} \cos(n\theta)\ J_n(\kmn r) \nonumber \\
\Theta_{mn}(r,\theta,t)&=&\Theta_0\ e^{\lmn t} \cos(n\theta)\ J_n(\kmn r)
\eqa
where $J_n$ is the $n$th Bessel function of the first kind ($\sin(n\theta)$ dependence is also acceptable).  The growth rates $\lmn$ and wave numbers $\kmn$ are determined via the boundary conditions, Eqs.~\ref{bcvel} and~\ref{bctempfix}.  Here with $w(r=1/2)=\Theta(r=1/2)=0$, $k_{m,n}$ is related to the $\rmn$ (the $m$th root of $J_n$) by 
\beq
\kmn=\frac{\rmn}{r_{ext}}=\frac{\rmn}{1/2}.
\eeq
The dispersion relation determining the growth rate for each mode is given by
\beq
2\sqrt{\Pra\,\ra}\ \lmn=-\kmn^2\left(1+\Pra\right)+ \sqrt{\kmn^4 \left(1+\Pra \right)^2+4\,\Pra\left( \ra-\kmn^4\right)},
\eeq
which has the same form as the tri-periodic case (up to a constant factor due to the different choices for non-dimensionalization), but the allowed values of $\kmn$ differ because of the geometry. 
The critical Rayleigh number for this case, i.e. for $\lambda_{1,1}(\ra_c)=0$, is $\ra_c = (k_{1 1})^4 = 3448.964$.

\begin{acknowledgements}
The authors acknowledge the National Computing Facilities (NCF) for awarding us computing time on SARA in Amsterdam. We thank Valentina Lavezzo for careful reading of the manuscript. 
We also thank the EU COST Action MP0806.
\end{acknowledgements}


\begin{thebibliography}{28}
\expandafter\ifx\csname natexlab\endcsname\relax\def\natexlab#1{#1}\fi

\bibitem[Ahlers {\em et~al.\/}(2009)Ahlers, Grossmann \& Lohse]{ahl09}
{\sc Ahlers, G., Grossmann, S. \& Lohse, D.} 2009 Heat transfer and large scale
  dynamics in turbulent {{{{Rayleigh-B\'enard}}}} convection. {\em Rev. Mod.
  Phys.\/} {\bf 81}, 503.

\bibitem[Calzavarini {\em et~al.\/}(2006)Calzavarini, Doering, Gibbon, Lohse,
  Tanabe \& Toschi]{cal06}
{\sc Calzavarini, E., Doering, C.~R., Gibbon, J.~D., Lohse, D., Tanabe, A. \&
  Toschi, F.} 2006 Exponentially growing solutions of homogeneous
  {{{{Rayleigh-B\'enard}}}} flow. {\em Phys. Rev. E\/} {\bf 73}, R035301.

\bibitem[Calzavarini {\em et~al.\/}(2005)Calzavarini, Lohse, Toschi \&
  Tripiccione]{cal05}
{\sc Calzavarini, E., Lohse, D., Toschi, F. \& Tripiccione, R.} 2005 Rayleigh
  and {{Prandtl}} number scaling in the bulk of {{{{Rayleigh-B\'enard}}}}
  turbulence. {\em Phys. Fluids\/} {\bf 17}, 055107.

\bibitem[Celani {\em et~al.\/}(2007)Celani, Mazzino, Seminara \& Tizzi]{cel07b}
{\sc Celani, A., Mazzino, A., Seminara, A. \& Tizzi, M.} 2007 Droplet
  condensation in two-dimensional bolgiano turbulence. {\em J. Turb.\/} {\bf
  8}, 1--9.

\bibitem[Chavanne {\em et~al.\/}(1997)Chavanne, Chilla, Castaing, Hebral,
  Chabaud \& Chaussy]{cha97}
{\sc Chavanne, X., Chilla, F., Castaing, B., Hebral, B., Chabaud, B. \&
  Chaussy, J.} 1997 Observation of the ultimate regime in
  {{{{Rayleigh-B\'enard}}}} convection. {\em Phys. Rev. Lett.\/} {\bf 79},
  3648--3651.

\bibitem[Chavanne {\em et~al.\/}(2001)Chavanne, Chilla, Chabaud, Castaing \&
  Hebral]{cha01}
{\sc Chavanne, X., Chilla, F., Chabaud, B., Castaing, B. \& Hebral, B.} 2001
  Turbulent {{{{Rayleigh-B\'enard}}}} convection in gaseous and liquid he. {\em
  Phys. Fluids\/} {\bf 13}, 1300--1320.

\bibitem[Cholemari \& Arakeri(2009)]{ara09}
{\sc Cholemari, M. \& Arakeri, J.} 2009 Axially homogeneous, zero mean flow
  buoyancy-driven turbulence in a vertical pipe. {\em J. Fluid Mech.\/} {\bf
  621}, 69--102.

\bibitem[Dubrulle(2001)]{dub01}
{\sc Dubrulle, B.} 2001 Momentum transport and torque scaling in
  {{Taylor-Couette}} flow from an analogy with turbulent convection. {\em Eur.
  Phys. J. B\/} {\bf 21}, 295.

\bibitem[Garaud {\em et~al.\/}(2010)Garaud, Ogilvie, Miller \&
  Stellmach]{gar10}
{\sc Garaud, P., Ogilvie, G., Miller, N. \& Stellmach, S.} 2010 A model of the
  entropy flux and reynolds stress in turbulent convection. {\em Month. Not.
  Roy. Astro. Soc.\/} {\bf 407}, 2451--2467.

\bibitem[Gibert {\em et~al.\/}(2006)Gibert, Pabiou, Chilla \& Castaing]{gib06}
{\sc Gibert, M., Pabiou, H., Chilla, F. \& Castaing, B.} 2006
  High-{{Rayleigh}}-number convection in a vertical channel. {\em Phys. Rev.
  Lett.\/} {\bf 96}, 084501.

\bibitem[Gibert {\em et~al.\/}(2009)Gibert, Pabiou, Tisserand, Gertjerenken,
  Castaing \& Chill\`{a}]{gib09}
{\sc Gibert, M., Pabiou, H., Tisserand, J.-C., Gertjerenken, B., Castaing, B.
  \& Chill\`{a}, F.} 2009 Heat convection in a vertical channel: Plumes versus
  turbulence diffusion. {\em Phys. Fluids\/} {\bf 21}, 035109.

\bibitem[van Gils {\em et~al.\/}(2011)van Gils, Huisman, Bruggert, Sun \&
  Lohse]{gil11}
{\sc van Gils, D., Huisman, S.~G., Bruggert, G.~W., Sun, C. \& Lohse, D.} 2011
  Torque scaling in turbulent {{Taylor-Couette}} flow with co- and
  counter-rotating cylinders. {\em Phys. Rev. Lett.\/} {\bf 106}, 024502.

\bibitem[Grossmann \& Lohse(2000)]{gro00}
{\sc Grossmann, S. \& Lohse, D.} 2000 Scaling in thermal convection: A unifying
  theory. {\em J. Fluid. Mech.\/} {\bf 407}, 27--56.

\bibitem[Grossmann \& Lohse(2001)]{gro01}
{\sc Grossmann, S. \& Lohse, D.} 2001 Thermal convection for large {{Prandtl}}
  number. {\em Phys. Rev. Lett.\/} {\bf 86}, 3316--3319.

\bibitem[Grossmann \& Lohse(2002)]{gro02}
{\sc Grossmann, S. \& Lohse, D.} 2002 Prandtl and {{Rayleigh}} number
  dependence of the {{Reynolds}} number in turbulent thermal convection. {\em
  Phys. Rev. E\/} {\bf 66}, 016305.

\bibitem[Grossmann \& Lohse(2011)]{gro11}
{\sc Grossmann, S. \& Lohse, D.} 2011 Multiple scaling in the ultimate regime
  of thermal convection. {\em Phys. Fluids\/} {\bf x}, y.

\bibitem[Kadanoff(2001)]{kad01}
{\sc Kadanoff, L.~P.} 2001 Turbulent heat flow: Structures and scaling. {\em
  Phys. Today\/} {\bf 54}~(8), 34--39.

\bibitem[Kim \& Moin(1985)]{kim85}
{\sc Kim, J. \& Moin, P.} 1985 Application of a fractional-step method to
  incompressible navier-stokes equations. {\em J. Comp. Phys.\/} {\bf 59},
  308--323.

\bibitem[Kraichnan(1962)]{kra62}
{\sc Kraichnan, R.~H.} 1962 Turbulent thermal convection at arbritrary
  {{Prandtl}} number. {\em Phys. Fluids\/} {\bf 5}, 1374--1389.

\bibitem[Landau \& Lifshitz(1987)]{ll87}
{\sc Landau, L.~D. \& Lifshitz, E.~M.} 1987 {\em Fluid Mechanics\/}. Oxford:
  Pergamon Press.

\bibitem[Lohse \& Toschi(2003)]{loh03}
{\sc Lohse, D. \& Toschi, F.} 2003 The ultimate state of thermal convection.
  {\em Phys. Rev. Lett.\/} {\bf 90}, 034502.

\bibitem[Lohse \& Xia(2010)]{loh10}
{\sc Lohse, D. \& Xia, K.-Q.} 2010 Small-scale properties of turbulent
  {{Rayleigh-B\'enard}} convection. {\em Ann. Rev. Fluid Mech.\/} {\bf 42},
  335--364.

\bibitem[Perrier {\em et~al.\/}(2002)Perrier, Morat \& LeMouel]{per02}
{\sc Perrier, F., Morat, P. \& LeMouel, J.~L.} 2002 Dynamics of air avalanches
  in the access pit of an underground quarry. {\em Phys. Rev. Lett.\/} {\bf
  89}, 134501.

\bibitem[Siggia(1994)]{sig94}
{\sc Siggia, E.~D.} 1994 High {{Rayleigh}} number convection. {\em Annu. Rev.
  Fluid Mech.\/} {\bf 26}, 137--168.

\bibitem[Spiegel(1971)]{spi71}
{\sc Spiegel, E.~A.} 1971 Convection in stars. {\em Ann. Rev. Astron.
  Astrophys.\/} {\bf 9}, 323--352.

\bibitem[Tisserand {\em et~al.\/}(2010)Tisserand, Creyssels, Gibert, Castaing
  \& Chill\`a]{tis10}
{\sc Tisserand, J.-C., Creyssels, M., Gibert, M., Castaing, D. \& Chill\`a, F.}
  2010 Convection in a vertical channel. {\em New J. Physics\/} {\bf 12},
  075024.

\bibitem[Verzicco \& Camussi(2003)]{ver03}
{\sc Verzicco, R. \& Camussi, R.} 2003 Numerical experiments on strongly
  turbulent thermal convection in a slender cylindrical cell. {\em J. Fluid
  Mech.\/} {\bf 477}, 19--49.

\bibitem[Verzicco \& Orlandi(1996)]{ver96}
{\sc Verzicco, R. \& Orlandi, P.} 1996 A finite-difference scheme for
  three-dimensional incompressible flow in cylindrical coordinates. {\em J.
  Comput. Phys.\/} {\bf 123}, 402--413.

\end{thebibliography}

\end{document}